\documentclass{article}

\usepackage{spconf}

\usepackage{url}
\usepackage{bs_macros}
\usepackage{float}
\usepackage[labelformat=simple]{subcaption}

\usepackage{tikz}
\usepackage{tikz-qtree}

\graphicspath{{./}}

\theoremstyle{definition}

\newtheorem{remark}{Remark}

\newtheorem{example}{Example}

\title{Portfolio Cuts: A Graph-Theoretic Framework to Diversification}
\name{Bruno Scalzo Dees $^{1}$, Ljubi$\check{\text{s}}$a Stankovi\'c $^{2}$, Anthony G. Constantinides $^{1}$, Danilo P. Mandic $^{1}$}

\address{$^{1}$Department of EEE, Imperial College London, London, SW7 2BT, UK \\
	$^{2}$Faculty of Electrical Engineering, University of Montenegro, Podgorica, 81000, Montenegro \\
	Emails: \{bs1912, a.constantinides, d.mandic\}@imperial.ac.uk, ljubisa@ucg.ac.me
}

\begin{document}
\ninept

\maketitle

\begin{abstract}
	\vspace{-0.2cm}
	Investment returns naturally reside on irregular domains, however, standard multivariate portfolio optimization methods are agnostic to data structure. To this end, we investigate ways for domain knowledge to be conveniently incorporated into the analysis, by means of \textit{graphs}. Next, to relax the assumption of the completeness of graph topology and to equip the graph model with practically relevant physical intuition, we introduce the \textit{portfolio cut} paradigm. Such a graph-theoretic portfolio partitioning technique is shown to allow the investor to devise robust and tractable asset allocation schemes, by virtue of a rigorous graph framework for considering smaller, computationally feasible, and economically meaningful clusters of assets, based on graph cuts. In turn, this makes it possible to fully utilize the asset returns covariance matrix for constructing the portfolio, even without the requirement for its inversion. The advantages of the proposed framework over traditional methods are demonstrated through numerical simulations based on real-world price data.
\end{abstract}

\begin{keywords}
	Financial signal processing, graph cut, graph signal processing, portfolio optimization, vertex clustering
\end{keywords}

\vspace{-0.2cm}

\section{Introduction}

\vspace{-0.2cm}

The introduction of \textit{modern portfolio theory} by Harry Markowitz in 1952 \cite{Markowitz1952} has marked the beginning of quantitative approaches to investing, with the underlying principle of \textit{diversification} becoming the cornerstone of decision-making in finance and economics. The theory suggests an optimal strategy for the investment, which is based on the first- and second-order moments of the asset returns. This optimization task is referred to as the \textit{mean-variance optimization} (MVO). Consider the vector, $\r(t) \in \domR^{N}$, which contains the returns of $N$ assets at a time $t$, the $i$-th entry of which is given by \vspace{-0.2cm}
\begin{equation}
	r_{i}(t) = \frac{p_{i}(t) - p_{i}(t-1)}{p_{i}(t-1)}
\end{equation} \vspace{-0.3cm}\\
where $p_{i}(t)$ denotes the value of the $i$-th asset at a time $t$. The MVO asserts that the optimal vector of asset holdings, $\w \in \domR^{N}$, is obtained through the following optimization problem \vspace{-0.1cm}
\begin{equation}
	\max_{\w} \;\{  \w^{\Trans}\boldmu - \lambda \w^{\Trans}\boldSigma\w \}
\end{equation} \vspace{-0.5cm}\\
where $\boldmu = \expect{\r} \in \domR^{N}$ is a vector of expected future returns, $\boldSigma = \cov{\r} \in \domR^{N \times N}$ is the covariance matrix of returns, and $\lambda$ is a Lagrange multiplier, also referred to as the \textit{risk aversion} parameter. In practice, it is usually necessary to impose additional constraints on the values of $\w$.

The growth of computational power has naturally made MVO a ubiquitous tool for financial practitioners, however, to date the validity of its underlying theory remains perhaps the most debated topic in the field. Among a number of issues that make MVO unreliable in practice, a major caveat is the well established sensitivity of MVO to perturbations of the estimates of $\boldmu$ and $\boldSigma$ \cite{Michaud1989,Michaud1998,Chopra1993}, whereby small changes in the inputs may generate portfolio holdings with vastly different compositions. This is largely because the inputs to the MVO are statistical estimates of the moments of non-stationary return distributions, which typically yield portfolios that are far from truly optimal ones; these may even exhibit poor performance and excessive turnover.


It has been empirically demonstrated that the key parameter, the expected returns $\boldmu$, can be rarely forecasted with sufficient accuracy. Consequently, various risk-based asset allocation approaches have been proposed, which drop the term $\boldmu$ altogether, with the optimization performed using $\boldSigma$ only. The most important example is the \textit{minimum variance} (MV) portfolio, formulated as \vspace{-0.1cm}
\begin{equation}
	\min_{\w} \;\;  \w^{\Trans}\boldSigma\w, \quad \text{s.t.} \;\; \w^{\Trans}\1 = 1
\end{equation} \vspace{-0.4cm}\\
where $\1=[1,...,1]^{\Trans}$, and the constraint, $\w^{\Trans}\1 = 1$, enforces full investment of the capital. The optimal portfolio holdings then become  \vspace{-0.2cm}
	\begin{equation}
		\w = \frac{\boldSigma^{-1}\1}{\1^{\Trans}\boldSigma^{-1}\1} \label{eq:minimum_variance}
	\end{equation} \vspace{-0.25cm}\\
However, even in the absence of $\boldmu$, the instability issues remain prominent, as the matrix inversion of $\boldSigma$ required in (\ref{eq:minimum_variance}) may lead to significant errors for ill-conditioned matrices.

\vspace{-0.2cm}

\begin{remark}
	The numerical instability issues associated with MV portfolio optimisation leads to a counter-intuitive result, whereby the more collinear the asset returns the greater the need for diversification, and the more unstable the portfolio solution as the inversion of matrices with collinear rows/columns is notoriously unstable  \cite{LopezdePrado2012,LopezdePrado2016}. Increasing the size of $\boldSigma$ further complicates the problem as more data samples are required to yield a positive-definite estimate, i.e. at least $\frac{1}{2}(N^{2}+N)$ independent and identically distributed (\textit{i.i.d.}) observations of $\r(t)$ are needed. The severe impact of these challenges is highlighted by the fact that, in practice, even naive (equally-weighted) portfolios, i.e. $\w = \frac{1}{N}\1$, have been shown to outperform the mean-variance and risk-based optimization solutions \cite{DeMiguel2009}.
\end{remark}

\vspace{-0.2cm}

These instability concerns have received substantial attention in recent years \cite{Kolm2014}, and alternative procedures have been proposed to promote robustness by either incorporating additional portfolio constraints \cite{Clarke2002}, introducing Bayesian priors \cite{Black1992} or improving the numerical stability of covariance matrix inversion \cite{Ledoit2003}. A more recent approach has been to model assets using \textit{market graphs} \cite{Boginski2003}, that is, based on graph-theoretic techniques. Intuitively, a universe of assets can naturally be modelled as a network of vertices on a graph, whereby an edge between two vertices (assets) designates both the existence of a link and the degree of similarity between assets \cite{Simon1962}. 

It is important to highlight that a graph-theoretic perspective offers an interpretable explanation for the underperformance of MVO techniques in practice. Namely, since the covariance matrix $\boldSigma$ is dense, standard multivariate models implicitly assume full connectivity of the graph, and are therefore not adequate to account for the structure inherent to real-world markets \cite{LopezdePrado2014,LopezdePrado2014_2,LopezdePrado2016}. Moreover, it can be shown that the optimal holdings under the MVO framework are inversely proportional to the vertex centrality, thereby over-investing in assets with low centrality \cite{Peralta2016,Li2019}. 

Intuitively, it would be highly desirable to remove unnecessary edges in order to more appropriately model the underlying structure between assets (graph vertices); this can be achieved through \textit{vertex clustering} of the market graph \cite{Boginski2003}. Various portfolio diversification frameworks employ this technique to allocate capital within and across clusters of assets at multiple hierarchical levels. For instance, the \textit{hierarchical risk parity} scheme \cite{LopezdePrado2016} employs an inverse-variance weighting allocation which is based on the number of assets within each asset cluster. Similarly, the \textit{hierarchical clustering based asset allocation} in \cite{Raffinot2017} finds a diversified weighting by distributing capital equally among each of the cluster hierarchies. 

Despite mathematical elegance and physical intuition, direct vertex clustering is an NP hard problem. Consequently, existing graph-theoretic portfolio constructions employ combinatorial optimization formulations \cite{Boginski2003,Boginski2005,Boginski2006,Gunawardena2012,Boginski2014,Kalyagin2014}, which too become computationally intractable for large graph systems. To alleviate this issue, we employ the \textit{minimum cut} vertex clustering method to introduce the \textit{portfolio cut}. In this way, smaller graph partitions (cuts) can be evaluated quasi-optimally, using algebraic methods, and in an efficient and rigorous manner. The proposed approach is shown to enable creation of graph-theoretic capital allocation schemes, based on measures of connectivity which are inherent to the portfolio cut formulation. Finally, it is shown that the proposed portfolio construction employs full information contained in the asset covariance matrix, and without requiring its inversion, even in the critical cases of limited data lengths or singular covariance matrices.

\vspace{-0.3cm}

\section{Portfolio Cuts}

\vspace{-0.2cm}


We follow the notation in \cite{Stankovic2019_1,Stankovic2019_2}, whereby a graph, $\mathcal{G} = \{\mathcal{V},\mathcal{E}\}$, is defined as a set of $N$ vertices, $\mathcal{V} = \{1,2,...,N\}$, which are connected by a set of edges, $\mathcal{E} \subset \mathcal{V} \times \mathcal{V}$. The existence of an edge between vertices $m$ and $n$ is designated by $(m, n) \in \mathcal{E}$. The strength of graph connectivity of an $N$-vertex graph can be represented by the \textit{weight matrix}, $\W \in \domR^{N \times N}$, with the entries defined as \vspace{-0.2cm}
\begin{equation}
	W_{mn} \begin{cases} 
		> 0, & (m,n) \in \mathcal{E},\\
		=0, & (m,n) \notin \mathcal{E},
	\end{cases}
\end{equation} \vspace{-0.35cm} \\
thus conveying information about the \textit{relative} importance of the vertex (asset) connections. The \textit{degree matrix}, $\D \in \domR^{N \times N}$, is a diagonal matrix with elements defined as \vspace{-0.2cm}
\begin{equation}
	D_{mm} = \sum_{n=1}^{N} W_{mn}
\end{equation} \vspace{-0.4cm} \\
and, and such, it quantifies the \textit{centrality} of each vertex in a graph. Another important descriptor of graph connectivity is the graph \textit{Laplacian matrix}, $\L \in \domR^{N \times N}$, defined as \vspace{-0.2cm}
\begin{equation}
	\L = \D - \W
\end{equation} \vspace{-0.6cm} \\
which serves as an operator for evaluating the curvature, or smoothness, of the graph topology.

\vspace{-0.3cm}


\subsection{Structure of market graph}

\vspace{-0.2cm}

A universe of $N$ assets can be represented as a set of vertices on a \textit{market graph} \cite{Boginski2003}, whereby the edge weight, $W_{mn}$, between vertices $m$ and $n$ is defined as the absolute correlation coefficient, $|\rho_{mn}|$, of their respective returns of assets $m$ and $n$, that is \vspace{-0.2cm}
\begin{equation}
	W_{mn} = \frac{|\sigma_{mn}|}{\sqrt{ \sigma_{mm}\sigma_{nn} }} = |\rho_{mn}|
\end{equation} \vspace{-0.4cm} \\
where $\sigma_{mn}=\cov{r_{m}(t),r_{n}(t)}$ is the covariance of returns between the assets $m$ and $n$. In this way, we have $W_{mn}=0$ if the assets $m$ and $n$ are statistically independent (not connected), and $W_{mn}>0$ if they are statistically dependent (connected on a graph). Note that the resulting weight matrix is symmetric, $\W^{\Trans}=\W$.

\pagebreak

\subsection{Minimum cut based vertex clustering}

Vertex clustering aims to group together vertices from the asset universe $\mathcal{V}$ into multiple disjoint \textit{clusters}, $\mathcal{V}_i$. For a market graph, assets which are grouped into a cluster, $\mathcal{V}_i$, are expected to exhibit a larger degree of mutual within-cluster statistical dependency than with the assets in other clusters, $\mathcal{V}_j$, $j\ne i$. The most popular classical graph cut methods are based on finding the minimum set of edges whose removal would disconnect a graph in some ``optimal'' sense; this is referred to as \textit{minimum cut} based clustering \cite{Schaeffer2007}. 

\begin{figure}[htp]
	\vspace{-0.3cm}
	\centering
	\includegraphics[scale=0.8,trim={7cm 21cm 7cm 2cm},clip]{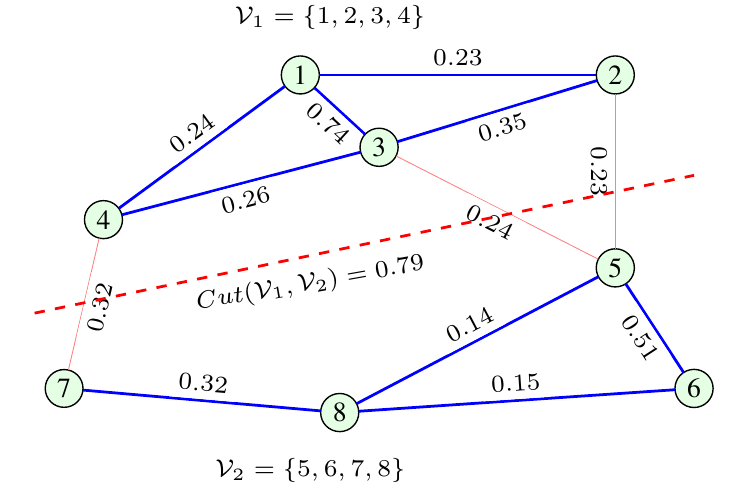}
	\vspace{-0.3cm}
	\caption{\footnotesize A cut for a graph with the disjoint subsets $\mathcal{V}_{1}=\{1,2,3,4\}$ and $\mathcal{V}_{2}=\{5,6,7,8\}$. The edges between the sets $\mathcal{V}_{1}$ and $\mathcal{V}_{2}$ are designated by thin red lines. The cut is equal to the sum of the weights that connect the sets $\mathcal{V}_{1}$ and $\mathcal{V}_{2}$, that is, $Cut(\mathcal{V}_{1},\mathcal{V}_{2})= 0.32+0.24+0.23=0.79$. }
	\label{fig:GSPb_ex2CUT}
	\vspace{-0.3cm}
\end{figure}

Consider an $N$-vertex market graph, $\mathcal{G} = \{\mathcal{V},\mathcal{E}\}$, which is grouped into $K=2$ disjoint subsets of vertices, $\mathcal{V}_{1} \subset \mathcal{V}$ and $\mathcal{V}_{2} \subset \mathcal{V}$, with  $\mathcal{V}_{1} \cup \mathcal{V}_{2}=\mathcal{V}$ and $\mathcal{V}_{1} \cap \mathcal{V}_{2}=\emptyset$. A cut of this graph, for the given clusters, $\mathcal{V}_{1}$ and $\mathcal{V}_{2}$, is equal to a sum of all weights that correspond to the edges which connect the vertices between the subsets, $\mathcal{V}_{1}$ and $\mathcal{V}_{2}$, that is 
\begin{equation}
	Cut(\mathcal{V}_{1},\mathcal{V}_{2})=\sum_{m \in \mathcal{V}_{1}} \sum_{n \in \mathcal{V}_{2} } W_{mn}  \label{eq:ideal_minimum_cut}
\end{equation} 
A cut which exhibits the minimum value of the sum of weights between the disjoint subsets, $\mathcal{V}_{1}$ and $\mathcal{V}_{2}$, considering all possible divisions of the set of vertices, $\mathcal{V}$, is referred to as \textit{the minimum cut}. Figure \ref{fig:GSPb_ex2CUT} provides an intuitive example of a graph cut.

Finding the minimum cut in (\ref{eq:ideal_minimum_cut}) is an NP-hard problem, whereby the number of combinations to split an even number of vertices, $N$, into \textit{any} two possible disjoint subsets is given by $C = 2^{(N-1)}-1$.

\begin{remark}
	To depict the computational burden associated with this brute force graph cut approach, even for typical market graph with $N = 500$ vertices (e.g. S\&P 500 stock index), the number of combinations to split the vertices into two subsets is $C = 1.6 \times 10^{150}$.
\end{remark}

\vspace{-0.1cm}

Within graph cuts, a number of optimization approaches may be employed to enforce some desired properties on graph clusters:

\smallskip

\noindent (i) \textit{Normalized minimum cut}. The value of $Cut(\mathcal{V}_{1},\mathcal{V}_{2})$ is regularised by an additional term to enforce the subsets, $\mathcal{V}_{1}$ and $\mathcal{V}_{2}$, to be \textit{simultaneously as large as possible}. The normalized cut formulation is given by \cite{Hagen1992} \vspace{-0.1cm}
\begin{equation}
CutN(\mathcal{V}_{1},\mathcal{V}_{2})=\Big(\frac{1}{N_{1}}+\frac{1}{N_{2}} \Big)\sum_{m \in \mathcal{V}_{1}} \sum_{n \in \mathcal{V}_{2} } W_{mn} \label{CutN}
\end{equation}  \vspace{-0.3cm} \\
where $N_{1}$ and $N_{2}$ are the respective numbers of vertices in the sets  $\mathcal{V}_{1}$ and $\mathcal{V}_{2}$. Since $N_{1}+N_{2}=N$, the term $\frac{1}{N_{1}}+\frac{1}{N_{2}}$ reaches its minimum for $N_{1}=N_{2}=\frac{N}{2}$.  

\pagebreak

\noindent (ii) \textit{Volume normalized minimum cut}. Since the vertex weights are involved when designing the size of subsets $\mathcal{V}_{1}$ and $\mathcal{V}_{2}$, then by defining \textit{the volumes} of these sets as $V_{1}=\sum_{n \in \mathcal{V}_{1}}D_{nn}$ and $V_{2}=\sum_{n \in \mathcal{V}_{2}}D_{nn}$, we arrive at \cite{Shi2000} \vspace{-0.2cm}
\begin{equation} 
CutV(\mathcal{V}_{1},\mathcal{V}_{2})=\Big(\frac{1}{V_{1}}+\frac{1}{V_{2}} \Big)\sum_{m \in \mathcal{V}_{1}} \sum_{n \in \mathcal{V}_{2} } W_{mn} \label{CutV}
\end{equation} \vspace{-0.3cm}\\
Since $V_{1}+V_{2}=V$, the term $\frac{1}{V_{1}}+\frac{1}{V_{2}}$ reaches its minimum for $V_{1}=V_{2}=\frac{V}{2}$. Notice that vertices with a higher degree, $D_{nn}$, are considered as structurally more important than those with lower degrees. In turn, for market graphs, assets with a higher average statistical dependence to other assets are considered as more \textit{central}.


\begin{remark}
	It is important to note that clustering results based on the two above graph cut forms are different. While the method (i) favours the clustering into subsets with (almost) equal number of vertices, the method (ii) favours subsets with (almost) equal volumes, that is, subsets with vertices exhibiting (almost) equal average statistical dependence to the other vertices.
\end{remark}

%

\subsection{Spectral bisection based minimum cut}

To overcome the computational burden of finding the normalized minimum cut, we employ an approximative spectral solution which clusters vertices using the eigenvectors of the graph Laplacian, $\L$. The algorithm employs the second (Fiedler \cite{Fiedler1973}) eigenvector of the graph Laplacian,  $\u_{2} \in \domR^{N}$,  to yield a \textit{quasi-optimal} vertex clustering on a graph. Despite its simplicity, the algorithm is typically accurate and gives a good approximation to the normalized cut \cite{Ng2002,Spielman2007}. 

To relate the problem of the minimum cut in (\ref{CutN}) and (\ref{CutV}) to that of eigenanalysis of graph Laplacian, we employ an \textit{indicator vector}, denoted by $\x \in \domR^{N}$ \cite{Stankovic2019_1}, for which the elements take sub-graph-wise constant values within each disjoint subset (cluster) of vertices, with these constants taking different values for different clusters of vertices. In other words, the elements of $\x$ uniquely reflect the assumed cut of the graph into disjoint subsets $\mathcal{V}_{1},\mathcal{V}_{2} \subset \mathcal{V}$. 

For a general graph, we consider two possible solutions for the indicator vector, $\x$, that satisfy the subset-wise constant form:

\smallskip

\noindent (i) \textit{Normalized minimum cut}. It can be shown that if the indicator vector is defined as \cite{Stankovic2019_1}
\begin{equation}
	x(n)=\begin{cases} 
		\frac{1}{N_{1}}, & \text{for } n \in \mathcal{V}_{1},  \\
		-\frac{1}{N_{2}}, & \text{for } n \in \mathcal{V}_{2},
	\end{cases} 
\end{equation}
then the normalized cut, $CutN(\mathcal{V}_{1}, \mathcal{V}_{2})$ in (\ref{CutN}), is equal to the Rayleigh quotient of $\L$ and $\x$, that is
\begin{equation}
	CutN(\mathcal{V}_{1}, \mathcal{V}_{2})=\frac{\x^{\Trans}\L\x}{\x^{\Trans}\x} \label{LqfCut}
\end{equation}
Therefore, the indicator vector, $\x$, which minimizes the normalized cut also minimizes (\ref{LqfCut}). This minimization problem, for the unit-norm form of the indicator vector,  can also be written as
\begin{equation}
\min_{\x} \;\; \x^{\Trans}\L\x, \quad \text{s.t.} \;\; \x^{\Trans}\x=1 \label{idicativeMIN}
\end{equation}
which can be solved through the eigenanalysis of $\L$, that is
\begin{equation}
	\L\x= \lambda_{k} \x
\end{equation} 
After neglecting the trivial solution $\x=\u_{1}$, ($k=1$), since it produces a constant eigenvector, we next arrive at $\x=\u_{2}$, ($k=2$).

\pagebreak

\noindent (ii) \textit{Volume normalized minimum cut}. Similarly, by defining $\x$ as
\begin{equation}
	x(n)=\begin{cases} 
		\frac{1}{V_{1}}, & \text{for } n \in \mathcal{V}_{1},  \\
		-\frac{1}{V_{2}}, & \text{for } n \in \mathcal{V}_{2},
	\end{cases} 
\end{equation}
the volume normalized cut, $CutV(\mathcal{V}_{1}, \mathcal{V}_{2})$ in (\ref{CutV}), takes the form of a generalised Rayleigh quotient of $\L$, given by \cite{Stankovic2019_1}
\begin{equation}
CutV(\mathcal{V}_{1}, \mathcal{V}_{2})=\frac{\x^{\Trans}\L\x}{\x^{\Trans}\D\x} \label{LqfCutV}
\end{equation} 
The minimization of (\ref{LqfCutV}) can be formulated as
\begin{equation}
\min_{\x} \;\; \x^{\Trans}\L\x, \quad \text{s.t.} \;\; \x^{\Trans}\D\x=1 \label{cutD}
\end{equation}
which reduces to a generalized eigenvalue problem of $\L$, given by
\begin{equation}
	\L\x =\lambda_{k} \D \x
\end{equation}
Therefore, the solution to (\ref{cutD}) becomes the generalized eigenvector of the graph Laplacian that corresponds to its lowest non-zero eigenvalue, that is $\x=\u_{2}$, ($k=2$).

\vspace{-0.1cm}

\begin{remark}
	The indicator vector, $\x$, converts the original, computationally intractable, combinatorial minimum cut problem into a manageable algebraic eigenvalue problem. However, the smoothest eigenvector, $\u_{2}$, of graph Laplacian is not subset-wise constant, and so such solution would be approximate but computationally feasible. 
\end{remark} 

\vspace{-0.1cm}

For the spectral solutions above, the membership of a vertex, $n$, to either the subset $\mathcal{V}_{1}$ or $\mathcal{V}_{2}$ is uniquely defined by the \textit{sign} of the indicator vector $\x=\u_{2}$, that is
\begin{equation}
	\mathrm{sign}(x(n))=\begin{cases} 
		1, & \text{for } n \in \mathcal{V}_{1},  \\
		-1, & \text{for } n \in \mathcal{V}_{2}.
	\end{cases} 
\end{equation}
Notice that a scaling of $\x$ by any constant would not influence the solution for clustering into subsets $\mathcal{V}_{1}$ or $\mathcal{V}_{2}$.

\vspace{-0.1cm}

\begin{remark} \label{remark:bottleneck}
	The value of the true normalized minimum cut in (\ref{CutN}) has been shown to be bounded from below and above with constants which are proportional to the smallest non-zero eigenvalue, $\u_{2}^{\Trans}\L\u_{2}=\lambda_{2}$, of the graph Laplacian \cite{Chung2005,Trevisan2013}. Therefore, the eigenvalue $\lambda_{2}$ serves as a measure of \textit{separability} of a graph, whereby the larger the value of $\lambda_{2}$, the less separable the graph.
\end{remark}

\vspace{-0.1cm}


\subsection{Repeated portfolio cuts}

Although the above analysis has focused on the case with $K=2$ disjoint sub-graphs, it can be straightforwardly generalized to $K \geq 2$ disjoint sub-graphs through the method of \textit{repeated bisection}. 

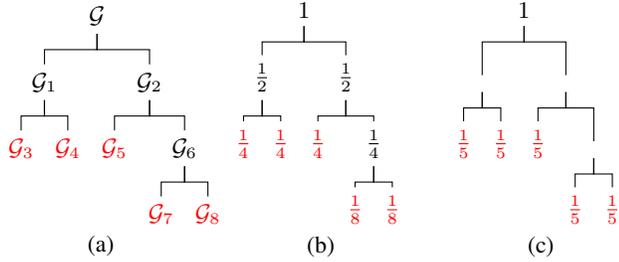
\begin{figure}[t]
	\vspace{-0.3cm}
	\centering
	\begin{subfigure}[t]{0.15\textwidth}
		\centering
		\begin{tikzpicture}
		\tikzset{level distance=25pt}
		\tikzset{edge from parent/.style=
			{draw,
				edge from parent path={(\tikzparentnode.south)
					-- +(0,-6pt)
					-| (\tikzchildnode)}}}
		\Tree [.$\mathcal{G}$ 
		[.$\mathcal{G}_{1}$ 
		[.\color{red}$\mathcal{G}_{3}$ ] 
		[.\color{red}$\mathcal{G}_{4}$ ] ]
		[.$\mathcal{G}_{2}$
		[.\color{red}$\mathcal{G}_{5}$ ] 
		[.$\mathcal{G}_{6}$ 
		[.\color{red}$\mathcal{G}_{7}$ ] 
		[.\color{red}$\mathcal{G}_{8}$ ] 
		] 
		] 
		]
		\end{tikzpicture}
		\vspace{-0.6cm}
		\caption{\label{fig:dendrogram}}
	\end{subfigure}
	\hfill
	\begin{subfigure}[t]{0.15\textwidth}
		\centering
		\begin{tikzpicture}
		\tikzset{level distance=25pt}
		\tikzset{edge from parent/.style=
			{draw,
				edge from parent path={(\tikzparentnode.south)
					-- +(0,-6pt)
					-| (\tikzchildnode)}}}
		\Tree [.$1$ 
		[.$\frac{1}{2}$ 
		[.\color{red}$\frac{1}{4}$  ] 
		[.\color{red}$\frac{1}{4}$  ] ]
		[.$\frac{1}{2}$ 
		[.\color{red}$\frac{1}{4}$  ] 
		[.$\frac{1}{4}$  
		[.\color{red}$\frac{1}{8}$  ] 
		[.\color{red}$\frac{1}{8}$  ] 
		] 
		] 
		]
		\end{tikzpicture}
		\vspace{-0.2cm}
		\caption{\label{fig:asset_allocation_i}}
	\end{subfigure}
	\hfill
	\begin{subfigure}[t]{0.15\textwidth}
		\centering
		\begin{tikzpicture}
		\tikzset{level distance=25pt}
		\tikzset{edge from parent/.style=
			{draw,
				edge from parent path={(\tikzparentnode.south)
					-- +(0,-6pt)
					-| (\tikzchildnode)}}}
		\Tree [.$1$ 
		[.{}
		[.\color{red}$\frac{1}{5}$  ] 
		[.\color{red}$\frac{1}{5}$  ] ]
		[.{} 
		[.\color{red}$\frac{1}{5}$  ] 
		[.{} 
		[.\color{red}$\frac{1}{5}$  ] 
		[.\color{red}$\frac{1}{5}$  ] 
		] 
		] 
		]
		\end{tikzpicture}
		\vspace{-0.2cm}
		\caption{\label{fig:asset_allocation_ii}}
	\end{subfigure}
	\vspace{-0.2cm}
	\caption{\footnotesize Graph asset allocation strategies. (a) Graph structure resulting from $K=4$ portfolio cuts. (b) $\frac{1}{2^{K_{i}}}$ scheme. (c) $\frac{1}{K+1}$ scheme.} 
	\vspace{-0.5cm}
\end{figure}

A single operation of the portfolio cut on the market graph, $\mathcal{G}$, produces two disjoint sub-graphs, $\mathcal{G}_{1}$ and $\mathcal{G}_{2}$, as illustrated in Figure \ref{fig:dendrogram}. Notice that in this way we construct a hierarchical binary tree structure, whereby the direct composition of the \textit{leaves} of the network is equal to the original market graph, $\mathcal{G}$. We can then perform a subsequent portfolio cut operation on one of the leaves based on some criterion (e.g. the leaf with the greatest number of vertices or volume). Therefore, $(K+1)$ disjoint sub-graphs (leaves) can be obtained by performing the portfolio cut procedure $K$ times. 

\begin{remark}
	Following Remark \ref{remark:bottleneck}, the maximum number of portfolio cuts, $K$, can be determined based on the value of the eigenvalue $\lambda_{2}$. For instance, the repeated portfolio cutting scheme may be terminated once the value of $\lambda_{2}$ exceeds a predefined threshold.
\end{remark}

\pagebreak

\begin{example}
	Figure \ref{fig:dendrogram} illustrates the hierarchical structure resulting from $K=4$ portfolio cuts of a market graph, $\mathcal{G}$. The leaves of the resulting binary tree are given by $\{\mathcal{G}_{3},\mathcal{G}_{4},\mathcal{G}_{5},\mathcal{G}_{7},\mathcal{G}_{8}\}$ (in red), whereby the number of disjoint sub-graphs is equal to $(K+1)=5$. Notice that the union of the leaves equals to the original graph, i.e.  $\mathcal{G}_{3} \cup \mathcal{G}_{4} \cup \mathcal{G}_{5} \cup \mathcal{G}_{7} \cup \mathcal{G}_{8} = \mathcal{G}$.
\end{example}


\vspace{-0.5cm}

\subsection{Graph asset allocation schemes}

\label{sec:asset_allocation}

\vspace{-0.2cm}

We next propose intuitive asset allocation strategies, inspired by the work in \cite{LopezdePrado2016,Raffinot2017}, which naturally builds upon the portfolio cut. The aim is to determine a diversified weighting scheme by distributing capital among the disjoint clusters (leaves) so that highly correlated assets within a given cluster receive the same total allocation, thereby being treated as a single uncorrelated entity.

By denoting the portion of the total capital allocated to a cluster $\mathcal{G}_{i}$ by $w_{i}$, we consider two simple asset allocation schemes:

\smallskip

\noindent (AS1) $w_{i} = \frac{1}{2^{K_{i}}}$, where $K_{i}$ is the number of portfolio cuts required to obtain sub-graph $\mathcal{G}_{i}$;

\smallskip

\noindent (AS2) $w_{i} = \frac{1}{K+1}$; where $(K+1)$ is the number of disjoint sub-graphs.

\smallskip

\vspace{-0.1cm}

\begin{remark}
	An equally-weighted asset allocation strategy may now be employed within each cluster, i.e. every asset within the $i$-th cluster, $\mathcal{G}_{i}$, will receive a weighting equal to $\frac{w_{i}}{N_{i}}$.
\end{remark}

\vspace{-0.2cm}

\begin{remark}
	The weighting scheme in AS1 above is closely related to the strategy proposed in \cite{Raffinot2017}, while the scheme in AS2 is inspired by the generic equal-weighted (EW) allocation scheme \cite{DeMiguel2009}. These schemes are convenient in that they require no assumptions regarding the across-cluster statistical dependence. In addition, unlike the EW scheme, they implicitly consider the inherent market risks (asset correlation) by virtue of the portfolio cut formulation, which is based on the eigenanalysis of the market graph Laplacian, $\L$.
\end{remark}

\vspace{-0.1cm}

\begin{example}
	Figures \ref{fig:asset_allocation_i} and \ref{fig:asset_allocation_ii} demonstrate respectively the asset allocation schemes in AS1 and AS2 for $K=4$ portfolio cuts, based on the market graph partitioning in Figure \ref{fig:dendrogram}. Notice that the weights associated to the disjoint sub-graphs (leaves in red) sum up to unity.
\end{example}

\vspace{-0.3cm}



\section{Numerical Example}

\enlargethispage{\baselineskip}

\vspace{-0.2cm}

The performance of the portfolio cuts and the associated graph-theoretic asset allocation schemes was investigated using historical price data comprising of the $100$ most liquid stocks in the S\&P 500 index, based on average trading volume, in the period 2014-01-01 to 2018-01-01. The data was split into: (i) the \textit{in-sample} dataset (2014-01-01 to 2015-12-31) which was used to estimate the asset correlation matrix and to compute the portfolio cuts; and (ii) the \textit{out-sample} (2016-01-01 to 2018-01-01), used to objectively quantify the profitability of the asset allocation strategies.

\begin{figure}[t]
	\vspace{-0.6cm}
	\centering
	\begin{subfigure}[t]{0.11\textwidth}
		\centering
		\includegraphics[width=1\textwidth, trim={0 0 0 0}, clip]{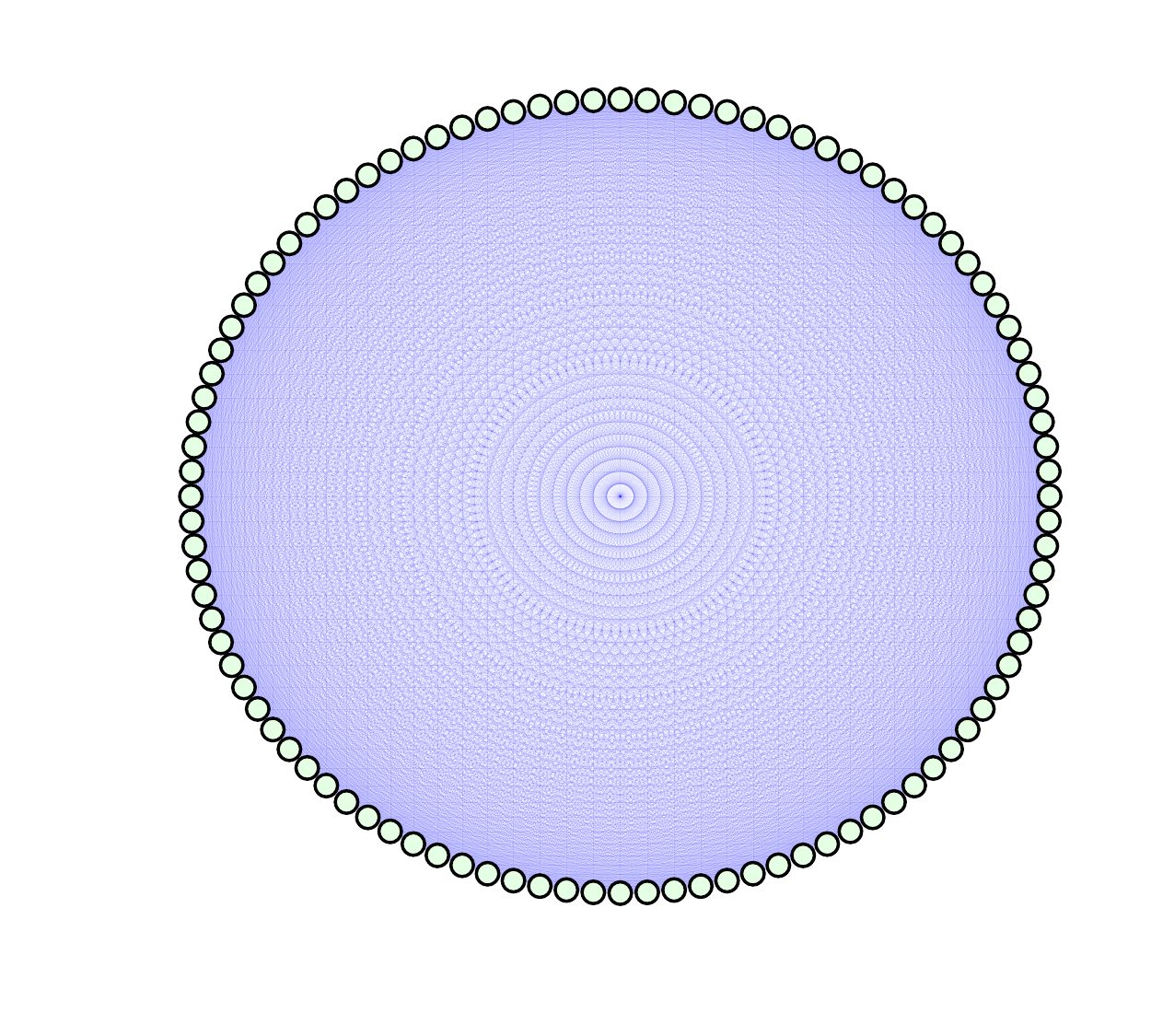} 
		\vspace{-0.7cm}
		\caption{\label{fig:fully_connected_graph}}
	\end{subfigure}
	\hfill
	\begin{subfigure}[t]{0.11\textwidth}
		\centering
		\includegraphics[width=1\textwidth, trim={0 0 0 0}, clip]{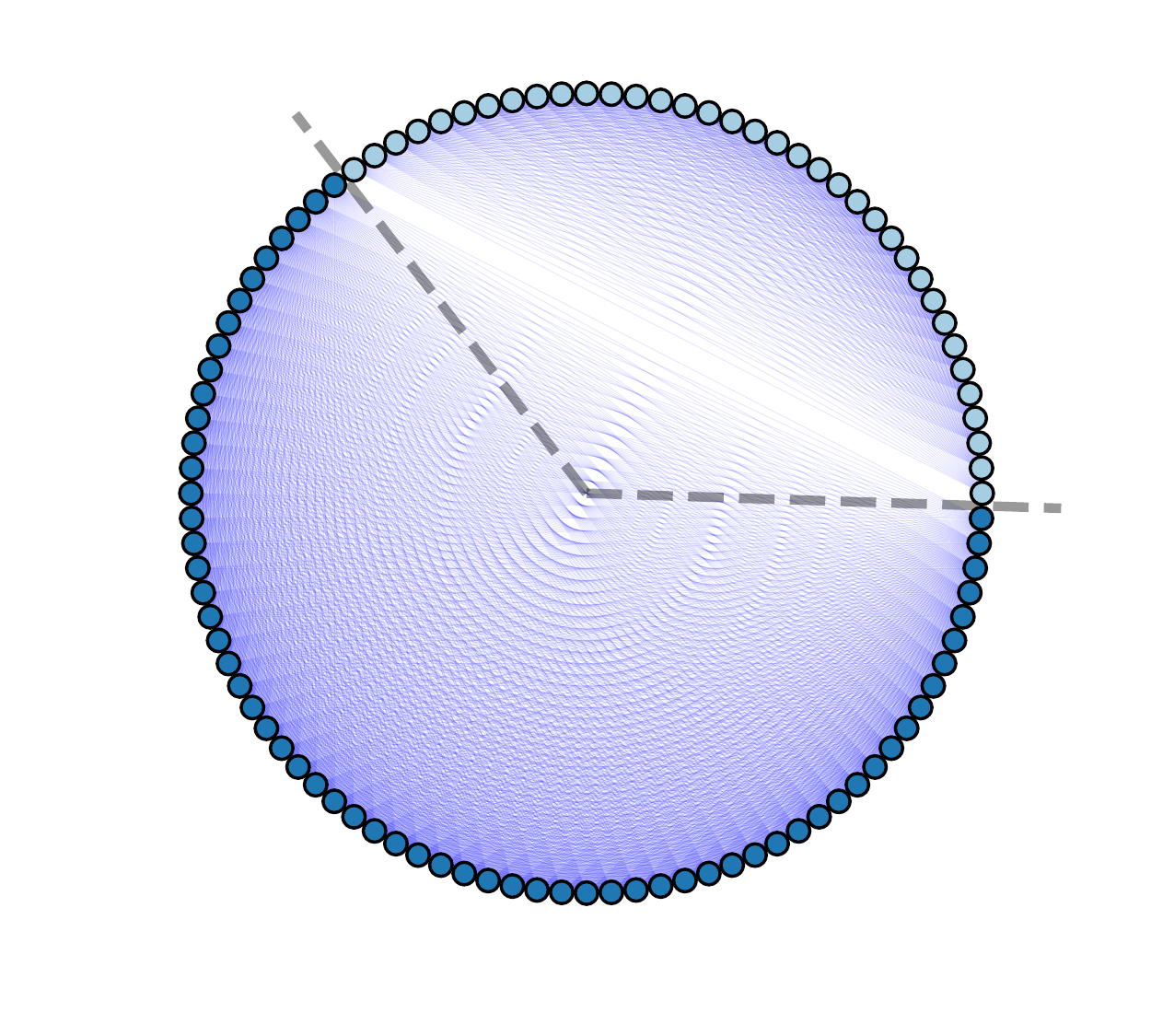} 
		\vspace{-0.7cm}
		\caption{}
	\end{subfigure}
	\hfill
	\begin{subfigure}[t]{0.11\textwidth}
		\centering
		\includegraphics[width=1\textwidth, trim={0 0 0 0}, clip]{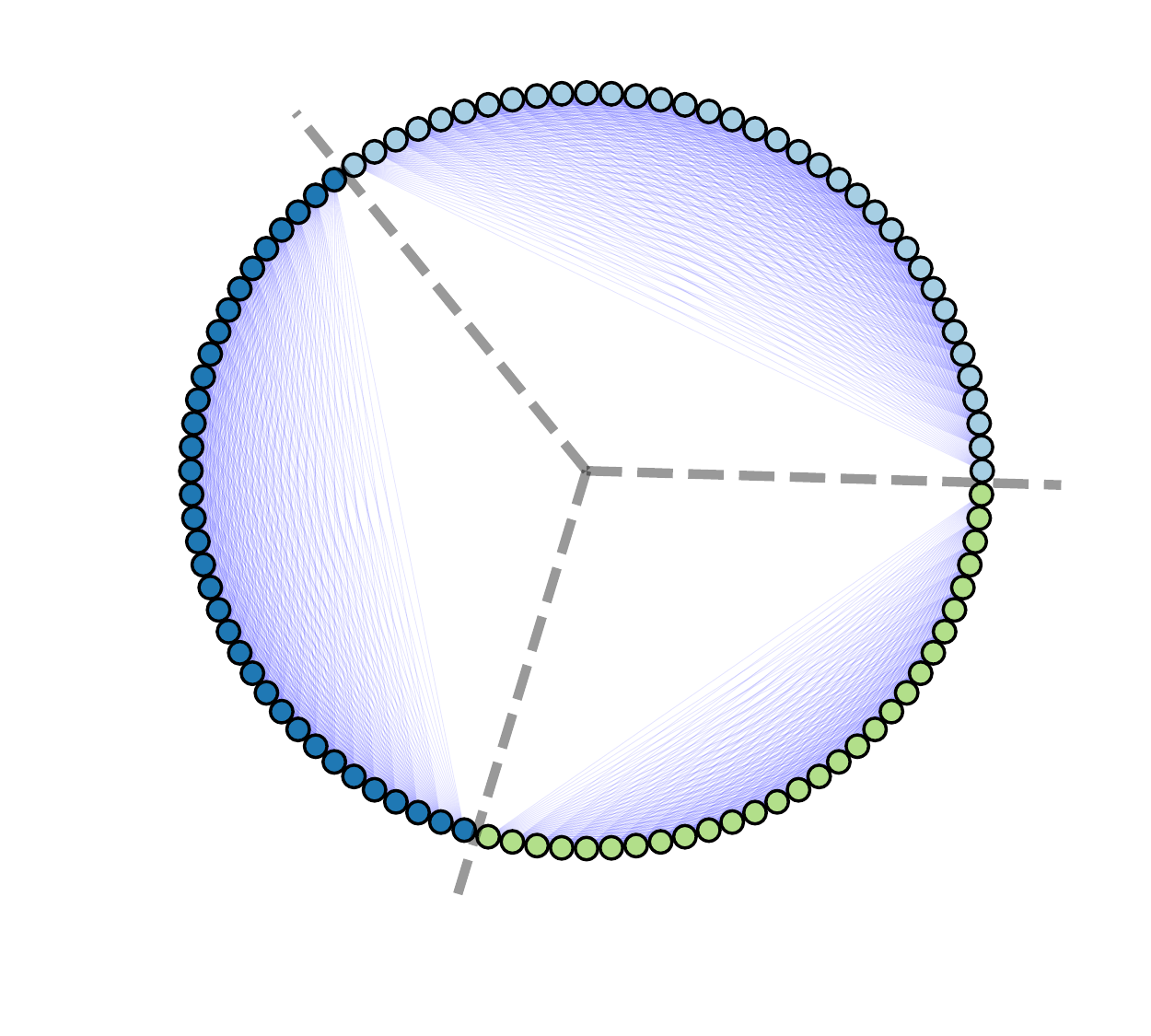} 
		\vspace{-0.7cm}
		\caption{}
	\end{subfigure}
	\hfill
	\begin{subfigure}[t]{0.11\textwidth}
		\centering
		\includegraphics[width=1.05\textwidth, trim={0 0 0 0.8cm}, clip]{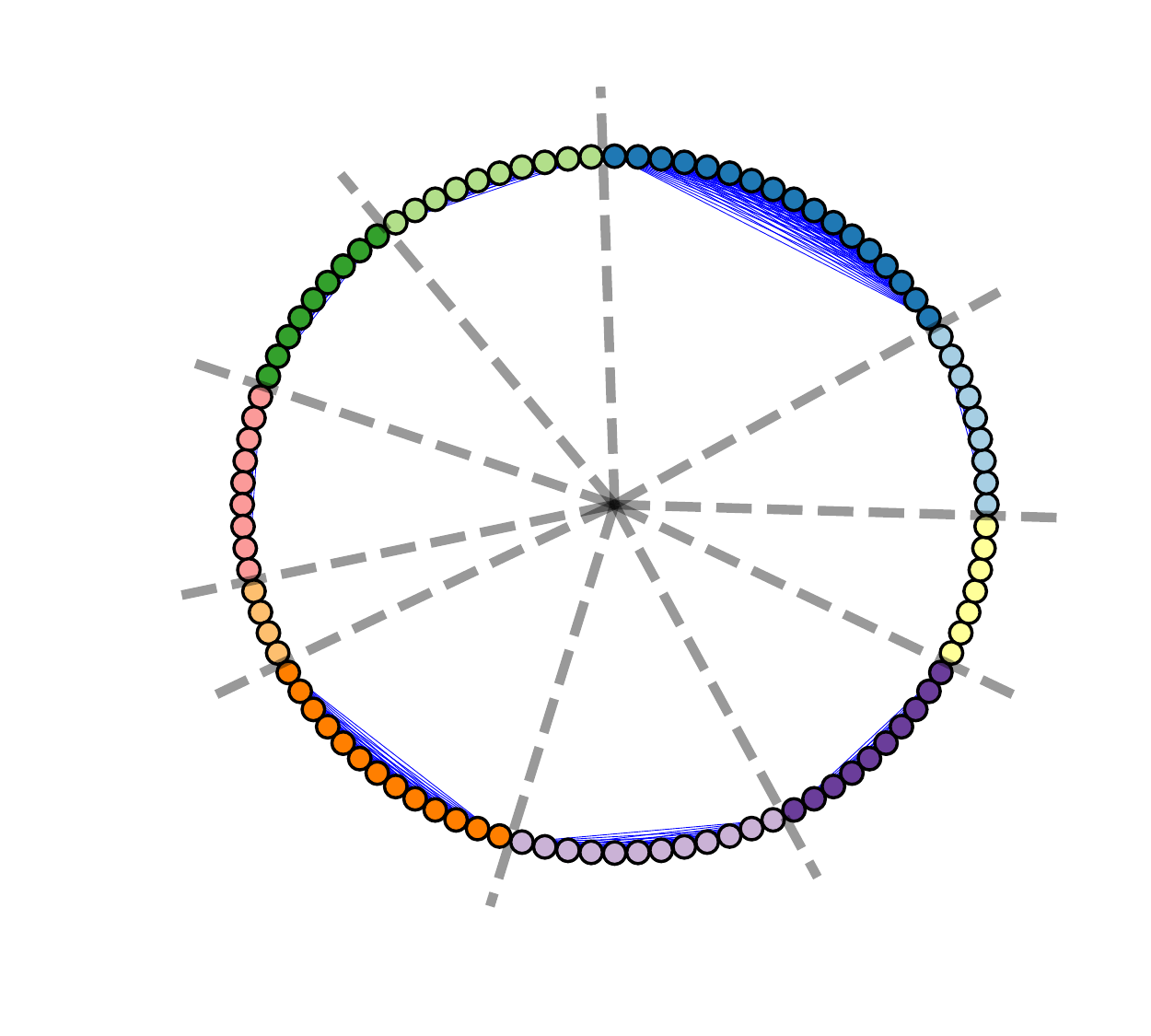} 
		\vspace{-0.7cm}
		\caption{\label{fig:connectivity_K10}}
	\end{subfigure}
	\vspace{-0.2cm}
	\caption{\footnotesize \label{fig:connectivity} Visualisation of the $100$-vertex market graph connectivity and its partitions into disjoint sub-graphs (separated by dashed grey lines). The edges (blue lines) were calculated based on the correlation between assets. (a) Fully connected market graph with $5050$ edges. (b) Partitioned graph after $K=1$ portfolio cuts (CutV), with $2746$ edges. (c) Partitioned graph after $K=2$ portfolio cuts (CutV), with $1731$ edges. (d) Partitioned graph after $K=10$ portfolio cuts (CutV), with $575$ edges. Notice that the number of edges required to model the market graph is significantly reduced with each subsequent portfolio cut, since $\sum_{i=1}^{K+1} \! \frac{1}{2}(N_{i}^{2} \! + \! N_{i}) < \frac{1}{2}(N^{2} \! + \! N)$, $\forall K>0$.} 
	\vspace{-0.7cm}
\end{figure}

Figure \ref{fig:connectivity} displays the $K$-th iterations of the proposed normalised portfolio cut in (\ref{LqfCut}), for $K=1,2,10$, applied to the original $100$-vertex market graph obtain from the in-sample data set.

\pagebreak

\begin{figure}[H]
	\vspace{-0.3cm}
	\centering
	\begin{subfigure}[t]{0.5\textwidth}
		\centering
		\includegraphics[width=1\textwidth, trim={0 0 0 0}, clip]{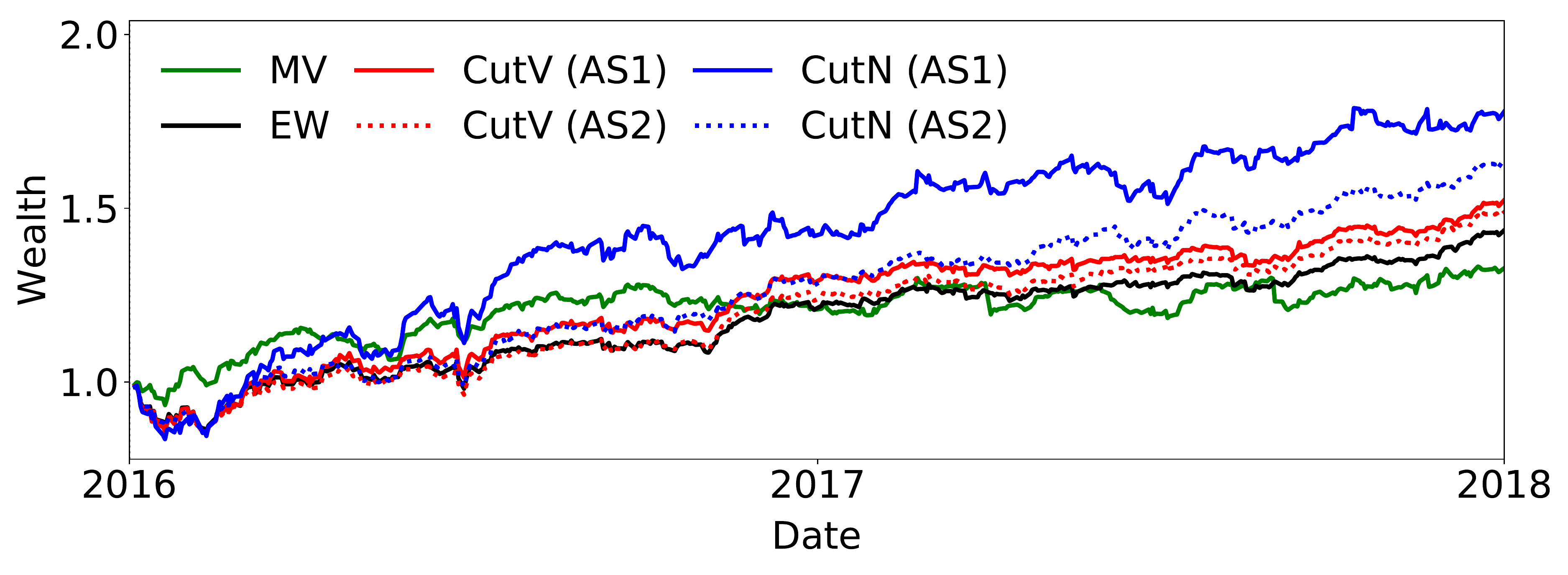} 
		\vspace{-0.6cm}
		\caption{Evolution of wealth for both the traditional (EW and MV) and graph-theoretic asset allocation strategies, based on ($K=10$) portfolio cuts.} 
		\vspace{0.2cm}
	\end{subfigure}
	\begin{subfigure}[t]{0.5\textwidth}
	\centering
	\scriptsize
	\renewcommand{\arraystretch}{1}
	\setlength{\tabcolsep}{3pt}
	\begin{tabular}[H!]{ c || c || c | c | c | c | c | c }
		\hline
		Cut Method & Allocation &  $K\!=\!1$ &  $K\!=\!2$ &  $K\!=\!3$ &  $K\!=\!4$ &  $K\!=\!5$ &  $K\!=\!10$ \\
		\hline
		CutV & AS1 &  $1.82$ &  $1.80$  &  $1.80$  &  $1.93$  &  $1.96$ &  $\boldsymbol{1.98}$  \\
		CutV & AS2 &  $1.82$ &  $1.81$ &  $1.94$ &  $2.03$ &  $1.95$ &  $\boldsymbol{2.05}$ \\
		CutN & AS1 &  $1.93$ &  $2.01$ &  $2.08$ &  $2.23$ &  $2.22$ &  $\boldsymbol{2.25}$ \\
		CutN & AS2 &  $1.93$ &  $2.04$ &  $2.17$ &  $\boldsymbol{2.65}$ &  $2.51$ &  $2.48$ \\
		\hline
	\end{tabular}
	\caption{Sharpe ratios attained for varying number of portfolio cuts $K$.}
	\vspace{-0.2cm}
	\end{subfigure}
	\caption{\footnotesize \label{fig:performance} Out-sample performance of the asset allocation strategies. Notice that the Sharpe ratio typically improves with each subsequent portfolio cut. The traditional portfolio strategies, EW and MV, attained the respective Sharpe ratios of $\text{SR}_{\text{EW}}=1.85$ and $\text{SR}_{\text{MV}}=1.6$.} 
	\label{fig:1}
	\vspace{-0.4cm}
\end{figure}

Next, for the out-sample dataset, graph representations of the portfolio, for the number of cuts $K$ varying in the range $[1,10]$, were employed to assess the performance of the asset allocation schemes described in Section \ref{sec:asset_allocation}. The standard equally-weighted (EW) and minimum-variance (MV) portfolios were also simulated for comparison purposes, with the results displayed in Figure \ref{fig:performance}. 

Conforming with the findings in \cite{LopezdePrado2016,Raffinot2017}, the proposed graph asset allocations schemes consistently delivered lower out-sample variance than the standard EW and MV portfolios, thereby attaining a higher \textit{Sharpe ratio}, i.e. the ratio of the mean to the standard deviation of portfolio returns. This verifies that the removal of possibly spurious statistical dependencies in the ``raw'' format, through the portfolio cuts, allows for robust and flexible portfolio constructions.

\section{Conclusions}

\enlargethispage{\baselineskip}

\vspace{-0.2cm}

A graph-theoretic approach to portfolio construction has been introduced which employs the proposed \textit{portfolio cut} paradigm to cluster assets using graph-specific techniques. The so derived graph asset allocation schemes have been shown to yield stable portfolio weights which are also robust to spurious asset correlations. Empirical analysis has demonstrated the advantages of the proposed framework over conventional portfolio optimization techniques, including a full utilization of the covariance matrix within the portfolio cut, without the requirement for its inversion. Finally, simulation results have demonstrated that the proposed framework allows for robust and flexible portfolio optimization, even in the critical cases of an ill-conditioned or singular asset covariance matrix.

\pagebreak


\bibliography{Bibliography}
\bibliographystyle{ieeetran}

\end{document}